\documentclass[runningheads]{llncs}
\usepackage{graphicx}
\graphicspath{ {./images/} }
\usepackage{multirow}
\usepackage{makecell}
\usepackage{subcaption}
\usepackage{siunitx}
\begingroup
\begin{document}
\title{Linear Prediction Residual for Efficient Diagnosis of Parkinson's Disease from Gait
}
\titlerunning{LP Residual for Parkinson's Diagnosis}
%
\author{
Shanmukh Alle\textsuperscript{1}\orcidID{0000-0001-5398-1154} 
\and
\\
U. Deva Priyakumar\textsuperscript{2}\orcidID{0000-0001-7114-3955} 
}
\authorrunning{Alle et al.}
\institute{Center for Computational Natural Sciences and Bioinformatics,
IIIT Hyderabad.
\email{shanmukh.alle@research.iiit.ac.in\textsuperscript{1}, deva@iiit.ac.in\textsuperscript{2}}}
%

%
%

\maketitle              
\begin{abstract}
Parkinson's Disease (PD) is a chronic and progressive neurological disorder that results in rigidity, tremors and postural instability. There is no definite medical test to diagnose PD and diagnosis is mostly a clinical exercise. Although guidelines exist, about 10-30\% of the patients are wrongly diagnosed with PD. Hence, there is a need for an accurate, unbiased and fast method for diagnosis. In this study, we propose LPGNet, a fast and accurate method to diagnose PD from gait. LPGNet uses Linear Prediction Residuals (LPR) to extract discriminating patterns from gait recordings and then uses a 1D convolution neural network with depth-wise separable convolutions to perform diagnosis. LPGNet achieves an AUC of 0.91 with a 21 times speedup and about 99\% lesser parameters in the model compared to the state of the art. We also undertake an analysis of various cross-validation strategies used in literature in PD diagnosis from gait and find that most methods are affected by some form of data leakage between various folds which leads to unnecessarily large models and inflated performance due to overfitting. The analysis clears the path for future works in correctly evaluating their methods.

\keywords{Parkinson's Diagnosis \and Gait \and Model Evaluation \and Convolutional Neural Networks \and Linear Prediction Analysis \and Signal Processing}
\end{abstract}

\section{Introduction}
Parkinson’s Disease (PD) is a neurological disorder that affects neurons in the brain responsible for motor control which leads to tremors, bradykinesia (slowed movement), limb rigidity, balance and gait problems. It is the second most common neurological disorder after Alzheimer's, affects about 10 million people worldwide\cite{dorsey2007projected} and is considered a chronic disease. Although the condition is not fatal, disease complications due to symptoms can be serious as they start gradually and develop over time. The Centers for Disease Control and Prevention (CDC) rates complications from PD as the 14\textsuperscript{th} largest cause of death in the United States of America\cite{flagg2021national}. Unfortunately, the cause of the condition is not yet known and there is no known cure for treating the condition. 

To this day there is no definite medical test to diagnose PD and diagnosis is still a clinical exercise\cite{massano2012clinical,berardelli2013efns} where an expert draws a conclusion from medical history, symptoms observed, and a neurological examination of a subject. Slow and gradual onset of symptoms and a possibility of human error make diagnosis inaccurate. About 10-30\% of the patients initially diagnosed with PD are later diagnosed differently\cite{berardelli2013efns,guardian_2019}. Although there is no known cure for treating the disorder, several therapies\cite{jankovic2012therapies,massano2012clinical} exist that have shown promise to improve the quality of life of affected patients and reduce severe complications. Development of a fast, standardized, and accurate method for diagnosis of PD is expected to help the lives of affected patients. 

Machine learning and statistical methods have shown promise in diagnosing various medical conditions in recent years. Several attempts have been made to use machine learning to build models to diagnose PD from various modalities like speech\cite{karabayir2020gradient,wroge2018parkinson}, handwriting patterns\cite{gil2019parkinson,thomas2017handwriting}, gait patterns\cite{lei2017joint,wahid2015classification}, etc. Although speech and handwriting based models perform well, they have a problem of large variability between demographics as they are not as universal as gait. Hence we believe that gait is the modality to look forward to, to build generalizable models for Parkinson's diagnosis. Few recent works using deep learning methods show good performance in diagnosis Parkinson's from gait. Zhao et al.\cite{zhao2018hybrid} use a hybrid CNN, LSTM model to achieve 98.6\% accuracy, Maachi et al.\cite{el2020deep} use a 1D CNN to achieve 98.7\% accuracy, Xia et al.\cite{xia2019dual} use a deep attention based neural network to achieve 99.07\% accuracy. Although all the works mentioned use some form of K-fold cross-validation to evaluate the performance of the model they build, there are a few significant differences in the methods used to generate cross-validation folds, making it difficult to compare performance reported in different studies.


We propose LPGNet (Linear Prediction residual based Gait classification Network) a deep learning based model that diagnoses PD from gait with good accuracy while being fast and small enough to be used in embedded systems to enable the method to be cheap and widely accessible. The secondary contribution is that we analyze and compare various methods for obtaining cross-validation folds (train-test splits) used in current literature to check them for data leakage and to gain clarity on how to correctly evaluate the model we build. The code for all the experiments discussed is made public to ensure reproducibility for future research\footnote[1]{\url{https://github.com/devalab/parkinsonsfromgait}}.

\section{Materials and Methods}
\subsection{Dataset}
We use a publicly available dataset\cite{PhysioNet} that is a collection of data collected from three different studies\cite{frenkel2005treadmill,yogev2005dual,hausdorff2007rhythmic}. The dataset consists of 306 gait recordings from 93 patients with PD and 73 healthy control subjects. Each recording is a two-minute long measurement of Vertical Ground Reaction Forces (VGRF) measured under each foot, sampled at a rate of 100Hz as a subject walks at their usual pace at ground level. Each recording includes 18 time series signals where 16 of them are VGRFs measured at 8 points under each foot and the remaining two represent the total VGRF under each foot. The database includes multiple recordings for some subjects where they were made to perform an additional task of solving arithmetic problems while walking.

\begin{figure}
     \centering
     \begin{subfigure}[b]{0.49\textwidth}
         \centering
         \includegraphics[width=\textwidth]{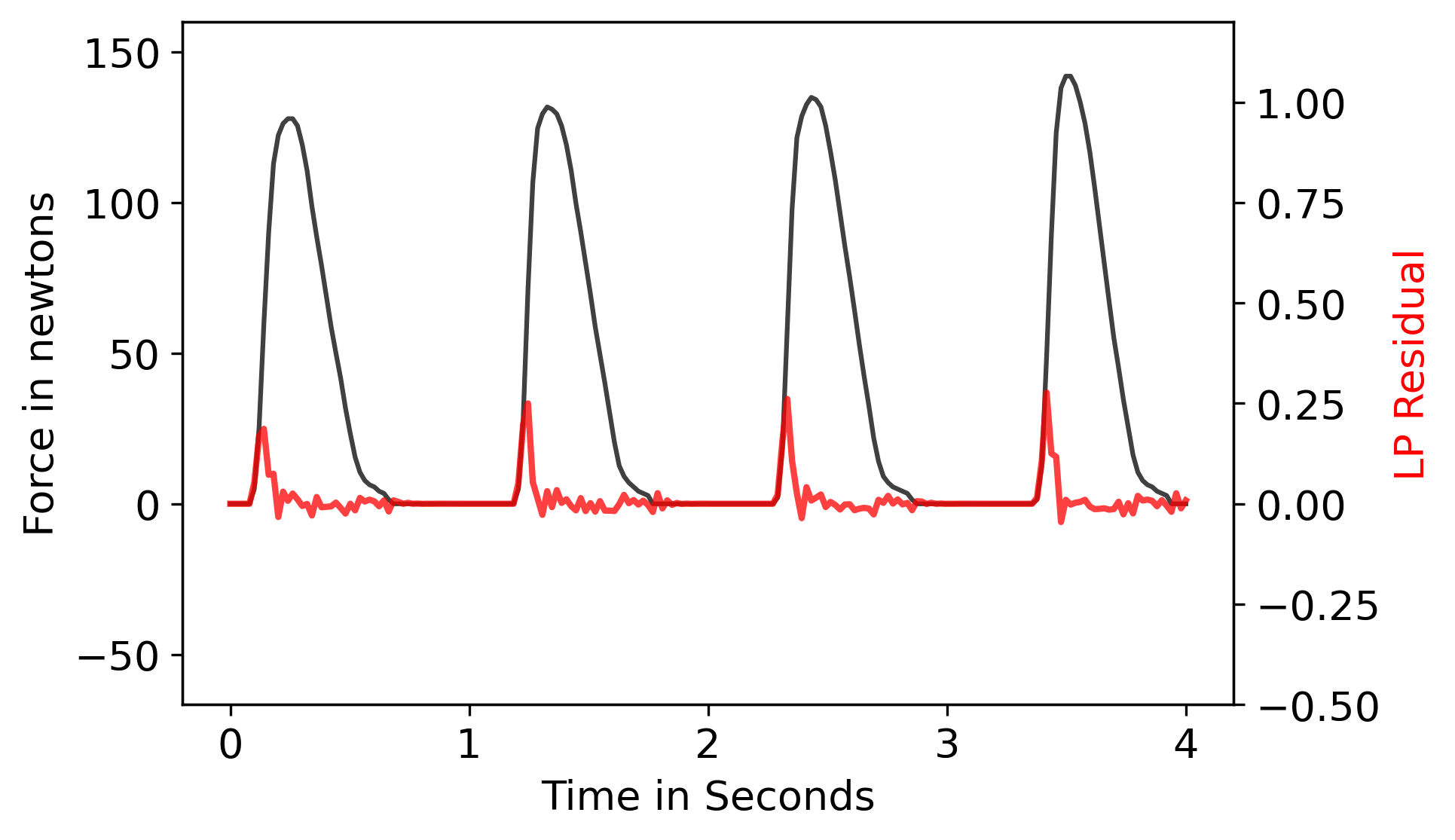}
         \caption{Normal Gait}
         \label{fig:normal gait}
     \end{subfigure}
     \hfill
     \begin{subfigure}[b]{0.49\textwidth}
         \centering
         \includegraphics[width=\textwidth]{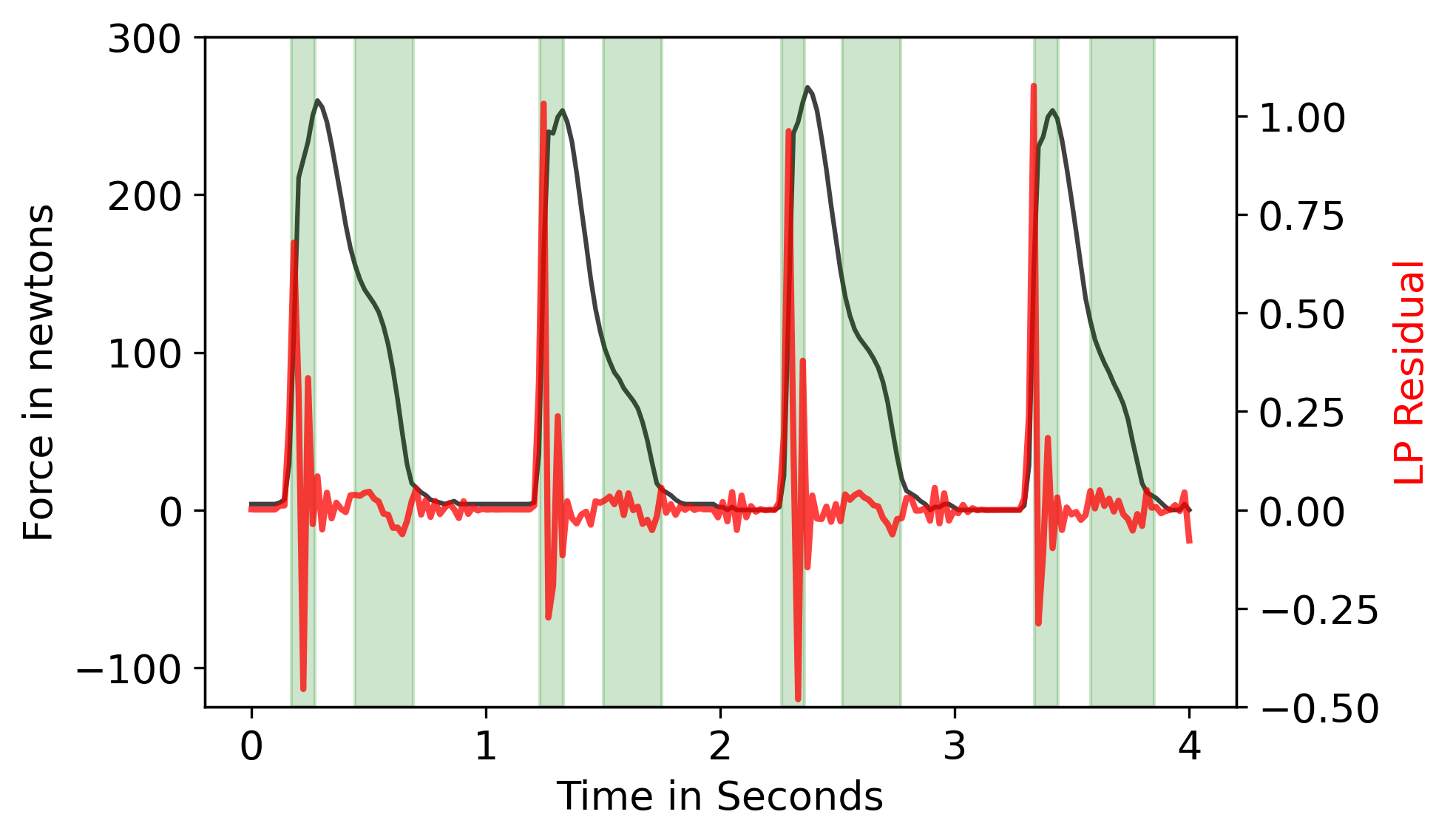}
         \caption{Parkinsonian Gait}
         \label{fig:parkinsonian gait}
     \end{subfigure}
        \caption{VGRFs measured from the sensor under the rear part of the right foot for a period of 4 seconds along with the aligned LPR along with anomalous regions highlighted in the case of the Parkinsonian gait recording.}
        \label{fig:gait Graphs}
\end{figure}

\subsection{Data-Leakage Experiment}
To evaluate different validation strategies mentioned in literature, we create a holdout test set that is used to measure performance of the best models obtained from each validation strategy to understand the presence of data leakage. The test set is made up of all VGRF recordings originating from 20\% of the subjects. A subject level separation is necessary as recordings originating from the same subject generally are similar. All the recordings are normalized to unit variance. Each recording is then split into windows of length 100 samples with a 50\% overlap. Each window inherits the class of the source recording and is considered a separate sample. We then split the windows from the remaining 80\% subjects available into the train and validation sets maintaining a 90:10 split with the following strategies. 
\begin{itemize}
    \item \textbf{Window Level:} Random 10\% of all the available windows make up the validation set and the remaining 90\% make up the train set. This method represents the validation strategies.\cite{xia2019dual,zhao2018hybrid,aversano2020early} 
    \item \textbf{Within Recording:} Random 10\% of windows from each recording make up the validation set and the remaining 90\% windows in each recording make up the train set. This method represents the validation strategy used by Maachi et al.\cite{el2020deep}
    \item \textbf{Subject Level:} Windows belonging to 10\% of the subjects make up the validation set and windows belonging to the remaining 90\% make up the train set.
\end{itemize}

We use a Convolution Neural Network (CNN) with three 1D convolution layers with RELU activation followed by a fully connected layer with sigmoid activation function to evaluate the differences between different validation strategies. The model is trained from scratch for each validation method with binary cross-entropy loss and L2 regularization. We use early stopping while training the model and choose the parameters from the epoch that gives the best validation accuracy to evaluate on the test set. We then examine the differences between the validation and test performance of the model representing each validation method to draw conclusions. Where ever necessary we use stratified splits to maintain the ratio of PD and control samples in the train, validation, and holdout test sets.
\begin{figure}
     \centering
    \includegraphics[width=\textwidth]{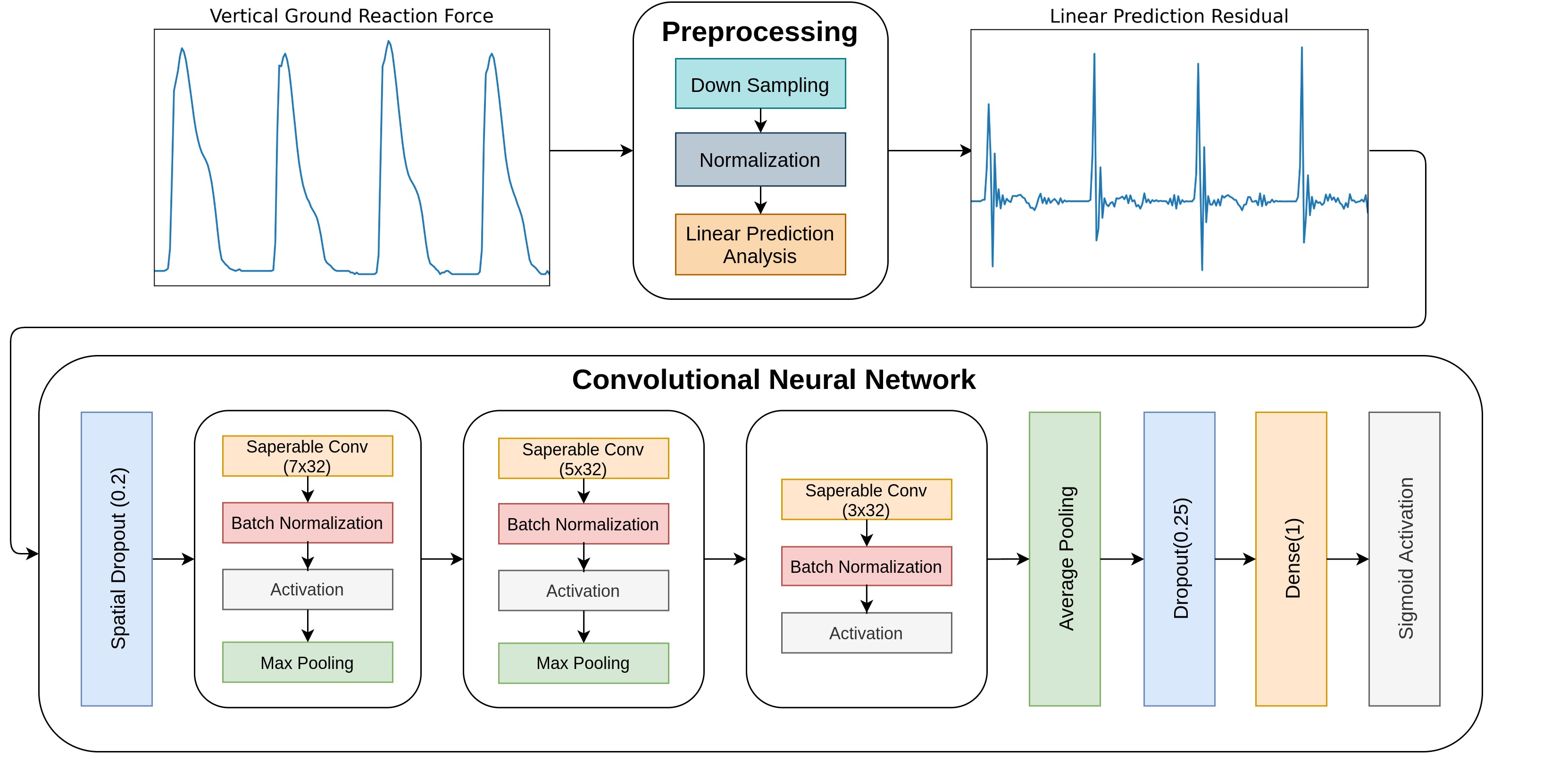}

        \caption{LPGNet (Linear Prediction residual Gait classification Network) pipeline for Parkinson's Diagnosis}
        \label{fig:modelpipeline}
\end{figure}

\subsection{Model Pipeline}
The prediction pipeline in LPGNet involves 3 main steps which include preprocessing, generating the LPR, and performing diagnosis with a CNN as shown in Figure \ref{fig:modelpipeline}. The following sections give further details.
\subsubsection{Preprocessing}
Each time series signal (VGRF measured at a point under the foot) is downsampled to 50 Hz and then normalized to unit variance. To avoid artifacts while downsampling, the raw signal is filtered with a moving average filter of order 2.
\subsubsection{Linear Prediction Residual:}Linear Prediction (LP) is a mathematical operation where future samples in a time series are estimated as a linear combination of $p$ past samples\cite{makhoul1975linear} as shown in equation \ref{eq:1}. The coefficients $a(i)$ of a linear predictor model information about the source of the time series. Linear Prediction Residual (LPR) is the prediction error $e(n)$ (equation \ref{eq:2}) that holds information specific to the time series that LP does not capture. This gives us the ability to separate normal gait patterns from the VGRF recordings and distill information specific to Parkinsonian gait in the residual.

\begin{equation}
\label{eq:1}
\widehat{x}(n)=-\sum_{i=1}^{p} a(i) x(n-i)
\end{equation}
\begin{equation}
\label{eq:2}
e(n)=x(n)-\widehat{x}(n)
\end{equation}
The coefficients of the linear predictor are obtained by minimizing the prediction error $e(n)$ for all samples in the signal which becomes the least-squares solution of a set of linear equations as mentioned in equation \ref{eq:3}. 
\begin{equation}
\label{eq:3}
X a=b
\end{equation}

\begin{equation}
X=\left[\begin{array}{cccc}
x(1) & 0 & \cdots & 0 \\
x(2) & x(1) & \cdots & 0 \\
\vdots & \vdots & \cdots & \vdots \\
x(p+1) & \cdots & \cdots & x(1) \\
\vdots & \vdots & \cdots & \vdots \\
0 & \cdots & 0 & x(m)
\end{array}\right], \quad a=\left[\begin{array}{c}
1 \\
a(1) \\
\vdots \\
a(p)
\end{array}\right], \quad b=\left[\begin{array}{c}
1 \\
0 \\
\vdots \\
0
\end{array}\right]
\end{equation}
\subsubsection{CNN Architecture:}
A 3-block 1D convolutional neural network with depth-wise separable convolutions is used as the clasifier in LPGNet as seen in Figure \ref{fig:modelpipeline}. In depth-wise separable convolutions\cite{DBLP:journals/corr/Chollet16a}, a normal convolution is replaced with channel-wise and point-wise convolutions in succession which reduces the computational and parametric complexity of the network while maintaining its predictive power. They are also less susceptible to overfitting compared to a normal CNN because of the lesser number of learnable parameters.  

Each convolution block contains a separable convolution layer, batch normalization layer, ELU activation followed by a max-pooling operation. The three blocks are followed by a global average pooling operation which computes the average over the time dimension resulting in a fixed dimensional vector for the final fully connected layer with sigmoid activation. Using global average pooling at this level enables the model to accept time series of varying lengths which in turn enables us to perform one step inference without resorting to padding/windowing. 
\subsubsection{Training:}
Training is done in two steps: Training the linear predictors to find optimal coefficients to generate the LPR followed by training a 1D-CNN to perform diagnosis. 

We train a separate linear predictor for each of the 18 time series in a gait recording. To find the LP coefficients, we concatenate all control gait recordings across time with padding equal to the order of the LP and then find the least-squares solution for each time series. Padding is added to prevent different recordings from affecting each other in the optimization process. Once we obtain the coefficients for the linear predictors, we generate LPRs for each of the 306 gait recordings which are then used to train a CNN. LPRs are used to train the CNN as normal gait characteristics in a subject's recording are removed when the predicted time series is subtracted from the original recording and the resulting LPR is more discriminating for classifying between normal and PD. We use the signal processing toolbox available in Matlab to generate LPRs.

Training the CNN is also done in two steps, the generated LPRs are divided into windows representing 2 seconds maintaining a 50\% overlap between successive windows. Each window is assigned the label of the source recording.  A CNN is then trained to classify these windows. Once the network converges keeping the weights of the convolution backbone of the network frozen, the fully connected layer is retrained to classify each two-minute recording at once. The second training step is necessary to calibrate the last fully connected layer to changes that might come up due to average pooling over the entire recording. The ADAM optimizer with its default parameters was used to train the network. Binary cross-entropy with label smoothing was used as the objective function.

Considering the small size of the dataset, different forms of regularization are used to control overfitting. We use spatial dropout\cite{DBLP:journals/corr/TompsonGJLB14} at the input to the CNN, followed by dropout at the input to the final logistic layer. L2 regularization on all learnable parameters along with gradient clipping was used to aid stability of the training process. We use TensorFlow library to implement and train the model. 

\subsection{Evaluation}
We use stratified 10-fold cross-validation while maintaining a subject level separation between the folds to evaluate the performance of the models considered. We report accuracy, AUC, and F1 scores with the mean and standard deviation measured over the 10 folds. We also report the number of trainable parameters and inference time to classify a two-minute recording on a single thread on an Intel Xeon E5-2640 v4 processor. The inference times reported are an average of 1000 runs. To enforce the use of a single thread we use the capabilities of the SLURM workload manager.  
    

\section{Results and Discussion}
Table \ref{table:Data leakage Experiment} summarizes the train, validation, and test performance of models achieving the best validation performance with each strategy considered (as seen in Section 2.2). We also report the absolute difference between each model’s performance on the validation and test sets to evaluate the degree of overfitting that in turn provides insight into the extent of data leakage in the validation strategy used. 
\begin{table}
\centering
    \caption{Data leakage experiment: Train, Validation and Test set performance of a baseline CNN for each validation strategy, expressed in the form of percentage accuracy (loss) to understand the presence of data leakage.}
    \label{table:Data leakage Experiment}
    \begin{tabular}{c c c c c}
    \hline
    Split Strategy & Train & Validation & Test & \makecell{Difference \\ (Validation-Test)}\\
    \hline
    Within Recording& \textbf{95.9 (0.285)}& \textbf{95.9 (0.284)}& 74.9 (0.637)& 21.0 (0.353)\\
    Window Level& 94.6 (0.301)& 94.1 (0.308)& 74.3 (0.661)& 19.8 (0.353)\\
    \textbf{Subject Level}& 88.7 (0.387)& 74.7 (0.572)& \textbf{78.8 (0.580)}& \textbf{4.1 (0.008)}\\
    \hline
    \end{tabular}
\end{table}

The window level and within recording split strategies show very good validation performance but perform poorly on the hold-out test set where the subject level split strategy performs the best. The within recording and window level split strategies show a large drop in performance between validation and test sets compared to the subject level split strategy. This signifies that these strategies are not good validation methods as it indicates heavy data leakage between the train and validation sets. Hence a subject level separation between the train and test folds should be maintained for correctly measuring the performance of a model when a holdout test set is not available. This explains the extremely good performance seen in works of Zhao et al.\cite{zhao2018hybrid}, Xia et al.\cite{xia2019dual}, Maachi et al.\cite{el2020deep} despite their models being relatively large for the number of training recordings available as large models overfit easily and generally perform very well in conditions where data leakage exists between the train and test sets.

\begin{table}[!h]
\centering
    \caption{Comparison of various baseline and ablation models}
    \label{table:Model Performances}
    \begin{tabular}{ c c c c c c}
    \hline
    Method & AUC & F1 Score &Accuracy& \makecell{Inference\\ Time (ms)} & Parameters\\
    \hline
    \textbf{LPGNet}&$\mathbf{91.7\pm9.4}$&$\mathbf{93.2\pm3.6}$&$\mathbf{90.3\pm5.8}$&\textbf{9.3ms}&4933\\
    Ablation&$90.4\pm8.1$&$91.2\pm4.9$&$87.6\pm6.7$&13.4ms&\textbf{4735}\\
    Baseline&$87.6\pm11.4$&$88.7\pm6.9$&$83.6\pm9.7$&20.6ms&16001\\
    1D-ConvNet\cite{el2020deep}&$86.7\pm10.3$&$88.2\pm6.8$&$82.5\pm10.1$&195.1ms&445841\\
    \hline
    \end{tabular}
\end{table}
As a baseline, we use the same model used in the data leakage experiment to represent the performance of a simple CNN. We average the probabilities predicted at the window level to get the prediction probabilities for a gait recording. We also compare with the 1D ConvNet model proposed by Maachi et al.\cite{el2020deep} when evaluated with a subject level 10 fold CV used in this work. To analyze the effect of the LPR we perform an ablation study where we train the proposed model with normalized VGRF signals. Table \ref{table:Model Performances} summarizes the performances of various methods considered in this study. The proposed LPGNet performs the best with an AUC of 91.7. It is also the fastest model with an inference time of 9.1 ms (3.7 ms for generating LPR and 5.6 ms for classification), the model is also considerably smaller than others with just 4,933 parameters (198 coefficients in 18 linear predictors of order 11 and 4,735 parameters in the 1D CNN). The large reduction in model size can be attributed to the corrected validation strategy used to evaluate the model while tuning the model architecture. We believe that the models built by Avasarano et al.\cite{aversano2020early} with 1.5 million parameters and Maachi et al.\cite{el2020deep} with 445,841 parameters are relatively large as their model validation strategies were biased towards large models due to data leakage between the train and validation folds. Additional reduction in model size and inference time can be attributed to the use of depth-wise separable convolutions that are more efficient. Faster inference speed can be attributed to the model’s ability to classify the entire sequence at once which removes the need for breaking the VGRF recordings into windows and averaging of window level predictions. 

\paragraph{Ablation Study:}A reduction in performance across the metrics was seen when using normalized VGRF signals at 100Hz to train the model. Apart from worse performance, an increase in variation (standard deviation) in accuracy and F1 score between folds is observed. This shows the role of the LPR in improving the performance and stability of the model. When using the LPR no loss in performance is observed when the VGRFs are sampled at 50Hz, this contributes to the faster inference compared to the ablation experiment despite LPR generation taking up additional time. 

LPR also provides a level of interpretability into how the model arrives at a decision as it is the error between modeled normal gait and real gait. A strong deviation from zero in the LPR signifies a deviation of the subject's gait from normal which indicates a higher chance of positive PD diagnosis. It can be seen in Figure \ref{fig:parkinsonian gait} that the deviations are higher in the case of Parkinsonian gait compared to normal gait. Since LPR has the same temporal resolution as the VGRF signal, we can identify parts of the gait cycle where a PD subject's gait deviates heavily from normal gait. The highlighted areas in the Parkinsonian gait in Figure \ref{fig:parkinsonian gait} point to such areas.

The ability of the proposed LPGNet to identify Parkinsonian gait accurately while being small and fast opens new avenues for it to be deployed in embedded systems with limited memory and compute resources which would go a long way in early diagnosis of Parkinson's in the developing countries of the world where clinical experts are not abundant. 

\section{Conclusion}
In this work, we present a novel method LPGNet that uses linear prediction residual with a 1D CNN to efficiently diagnose Parkinson’s from VGRF recordings. The proposed method achieves good discriminative performance with an accuracy of 90.3\% and an F1 score of 93.2\%. The model proposed is also orders of magnitude smaller and faster than methods described in literature. The proposed linear prediction residual aids in improving the interpretability of the method by pointing to the positions in gait patterns that deviate from normal. We also evaluate different validation strategies used in literature and identify the presence of data leakage and show that a subject level separation is necessary for correct evaluation of a method. This experiment clears the path for future works in correctly evaluating their methods by identifying sub-optimal strategies that are susceptible to data leakage.  

\paragraph{{\normalfont \textbf{Acknowledgements.}}} This study was supported by funding from IHub-Data and IIIT Hyderabad. We would also like to thank Dr. K Sudarsana Reddy for the discussions we had regarding the theoretical correctness of the method presented.
\bibliographystyle{splncs04}
\bibliography{references}
\end{document}


\title{Linear Prediction Residual for Efficient Diagnosis of Parkinson's Disease from Gait: Supplementary Document}
\titlerunning{LP Residual for Parkinson's Diagnosis}
%
\author{
Shanmukh Alle\textsuperscript{1}\orcidID{0000-0001-5398-1154} 
\and
\\
U. Deva Priyakumar\textsuperscript{2}\orcidID{0000-0001-7114-3955} 
}
\authorrunning{Alle et al.}
\institute{Center for Computational Natural Sciences and Bioinformatics,
IIIT Hyderabad.
\email{shanmukh.alle@research.iiit.ac.in\textsuperscript{1}, deva@iiit.ac.in\textsuperscript{2}}}

\maketitle
\subsection*{Baseline CNN Architecture}
\begin{figure}
     \centering
    \includegraphics[width=\textwidth]{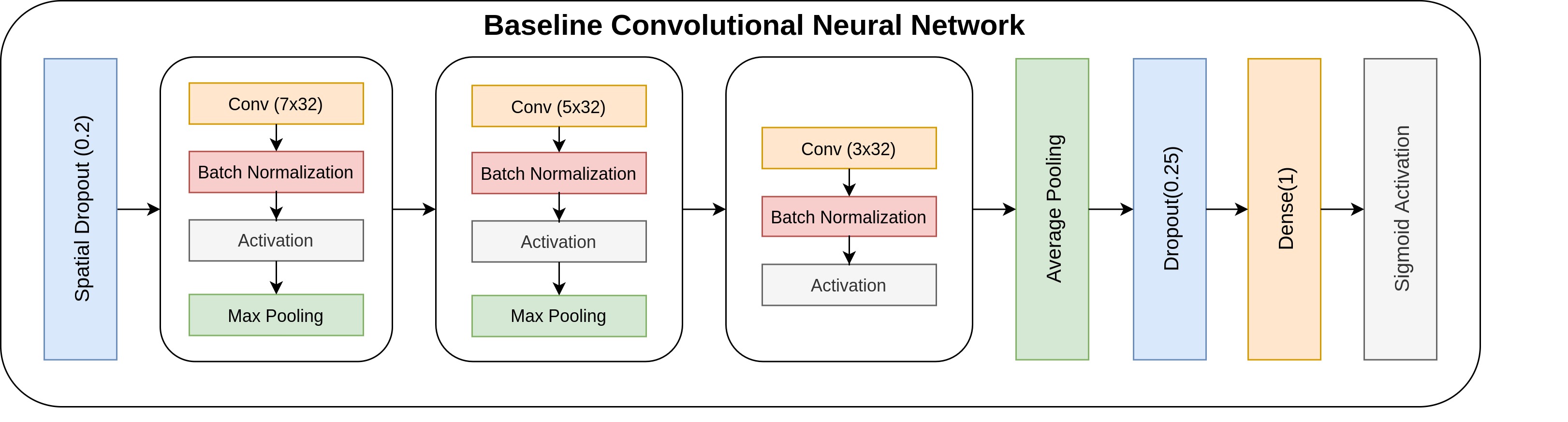}

        \caption{Baseline CNN Architecture}
        \label{fig:modelpipeline}
\end{figure}

\subsection*{Hyperparameters Used}
\paragraph{LPGNet:}
The ADAM optimizer with its default parameters was used to train the network. Binary cross entropy with label smoothing of 0.1 was used as the objective function. A batch size of 128 and learning rate of \num{5e-4} is used in first training step and a batch size of 64 and learning rate of \num{1e-3} is used in the second training step. The learning rate was reduced by a factor of \num{4} when a plateau was observed. 
\paragraph{Baseline:}
The ADAM optimizer with its default parameters was used to train the network. Binary cross entropy with label smoothing of 0.1 was used as the objective function. A batch size of 800 and learning rate of \num{1e-3} was used to train the model. The learning rate was reduced by a factor of \num{4} when a plateau was observed. 

\subsection*{Hardware used for running Experiments}
All experiments were conducted on a machine with dual Intel Xeon E5-2640 v4 processors and four GTX-1080ti GPUs